\documentclass[useAMS,usenatbib]{mn2e}
\usepackage{graphicx}% Include figure files
\usepackage[bookmarks]{hyperref}
\usepackage{amsmath}
\usepackage{amssymb}
\usepackage{color}
\title[Stellar feedback from HMXBs in cosmological hydrodynamical simulations]{Stellar feedback
  from high-mass X-ray binaries in cosmological hydrodynamical simulations}
\author[M. C. Artale, P. B. Tissera \& L. J. Pellizza]{M. C. Artale$^{1}$\thanks{E-mail:
mcartale@iafe.uba.ar} ,  P. B. Tissera$^{1,2,3}$ \& L. J. Pellizza$^{4}$ \\
$^{1}$Instituto de Astronom\'ia y F\'isica del Espacio (IAFE,
CONICET-UBA), C.C. 67 Suc. 28, C1428ZAA Ciudad de Buenos Aires, Argentina.\\
$^{2}$Departamento de Ciencias Fisicas, Universidad Andres Bello,
Av. Republica 220, Santiago, Chile. \\
$^{3}$Millennium Institute of Astrophysics, Av. Republica 220, Santiago, Chile. \\
$^{4}$Instituto Argentino de Radioastronom\'ia, CONICET, Camino Gral. Belgrano km 40, Berazategui, Prov. de Buenos Aires, Argentina \\
}
\begin{document}

%\date{Accepted 1988 December 15. Received 1988 December 14; in original form 1988 October 11}

\pagerange{\pageref{firstpage}--\pageref{lastpage}} \pubyear{2014}

\maketitle

\label{firstpage}

\begin{abstract}
We explored the role of X-ray binaries composed by a black hole and a
massive stellar companion  (BHXs) as sources of kinetic feedback by using 
hydrodynamical cosmological simulations. 
 Following previous results, our BHX model selects
 low metal-poor stars ($Z = [0,10^{-4}]$)  as possible progenitors. The model that better reproduces observations
  assumes that a $\sim 20\%$ fraction of low-metallicity black holes are in
  binary systems which produce BHXs. These sources are estimated to 
  deposit $\sim 10^{52}$ erg of kinetic energy per event. With
  these parameters and in the simulated volume, we find that the energy
injected by BHXs represents $\sim 30\%$ of the total energy released
by SNII and BHX events at redshift $z\sim7$ and then decreases rapidly as baryons get chemically enriched. 
Haloes with virial masses  smaller  than $\sim 10^{10} \,
M_{\odot}$ (or $T_{\rm vir} \lesssim 10^5 $ K) are the most directly
affected ones by BHX feedback. These haloes host galaxies with stellar masses
in the range $10^7 - 10^8$ M$_\odot$.
Our results show that BHX feedback is able to  keep the interstellar
medium warm,
without removing a significant gas fraction, in agreement with previous
analytical calculations. Consequently,  the
stellar-to-dark matter mass ratio is better reproduced at high redshift.
 Our model also predicts a stronger evolution of the number of galaxies as a
function of the stellar mass with redshift
when BHX feedback is considered. These findings support previous claims that the BHXs could be an effective source
of feedback in early stages of galaxy evolution.

\end{abstract}


\section{Introduction}

The regulation of star formation activity  (SF) in galaxies of
different masses  is still an open problem in cosmological
simulations.
Different feedback mechanisms have been proposed to regulate the
transformation of gas into stars.  Among them, supernova (SN) feedback is considered to play a
critical role in the regulation of SF, with a larger impact for galaxies with circular
velocities lower than $\sim 100 \, {\rm km\, s}^{-1}$ \citep{dekelsilk1986}.
The SN feedback modelling in
hydrodynamical codes has certainly helped to reach better agreement of
the properties of the simulated galaxies with
observations \citep[e.g.][]{Governato2007,springel2006,Scannapieco2006}.
However, these studies also show that SN feedback might  not be efficient enough to regulate
the SF activity in low-mass systems,
principally at high redshift. This problem manifests itself, for example, in
the apparent inconsistency between the stellar-to-virial mass relation obtained
from abundance matching techniques and that of hydrodynamical
simulations in the low-mass end
\citep[e.g.][]{Sawala2011,Moster2013,Behroozi2013}. Different
feedback mechanisms such as cosmic rays, radiative feedback or
photoionization, among others, have been  explored with the aim of
improving the regulation of the SF in numerical models
\citep[e.g.][]{Wadepuhl2011,Stinson2012,Hopkins2013,Ceverino2014}. However, much
work is still needed to both understand the physical processes which regulate
star formation as a function of  redshift and  to improve their modelling in numerical simulations.

High-mass X-ray binaries (HMXBs) are accretion-powered stellar systems
composed by a black hole or a neutron star and a companion massive star, which
emit X rays at typical luminosities of $10^{38}\,{\rm erg \, s}^{-1}$
\citep[see][for a review]{Fabbiano2006} and hence provide a radiative feedback
on the interstellar medium. In some cases, they also produce collimated outflows
(jets) with similar kinetic luminosities. These jets act as another feedback 
mechanism, heating the surrounding medium. The interest on
HMBXs has increased in the last years because theoretical and observational
works suggest that the production, and possibly the luminosity of these
sources, could increase with decreasing metallicity of the stellar progenitor
\citep[e.g.][]{Dray2006,Linden2010,Kaaret2011,Brorby2014}. Under this
hypothesis, HMBXs could play an important role in the formation and evolution
of galaxies at early epochs of the Universe.

Given their radiative feedback, HMXBs have been proposed as effective
sources of reionization of the intergalactic medium (IGM) and of heating of the
gas component in the small haloes in the very early Universe
\citep[e.g.][]{Power2009}. A recent report from \citet{Fragos2013B} use the
results of combining semi-analytical models of galaxy formation with a
population synthesis code \citep{Fragos2013A}, to compare the X-ray luminosity
density produced by HMXBs and active galactic nuclei as a function of redshift,
claiming that the first prevails at $z \gtrsim 6-8$. Different works have also
studied the role of these binaries in the heating of the IGM and the shaping of
the 21~cm signal from first galaxies
\citep{Power2013,Kaaret2014,Fialkov2014,Pacucci2014}. HMXBs have also been
investigated as a source of reionization. \citet{Mirabel2011} suggest that the
rate of ionizing photons emitted by HMXBs with black holes might be greater
than that of their progenitor stars, and that these sources could heat the IGM
up to $10^4$~K, keeping it ionized. Using zoomed hydrodynamical simulations of
mini-haloes, \citet{Jeon2013} have analysed the feedback from Population~III
(hereafter Pop~III) HMXBs. Their results suggest that X-ray photons from these
sources may suppress small scale structures and reduce the recombination rate
in the IGM, providing a net positive feedback on reionization.
\citet{Knevitt2014} investigated the effect of HMXBs on the high redshift IGM
using a  one-dimensional radiative transfer code. Contrary to previous works,
they found that HMXBs do not produce neither any significant additional
ionization nor heating of the IGM, except for the distant IGM in the case of
continuous star formation. As it can be seen from the above discussion, the
issue of the effects of the radiative feedback from HMXBs is far from being
settled.

Recently, \citet{Justham2012} studied high-mass X-ray binaries (HMXBs) as a
potential source of feedback through the kinetic energy of their jets.
Those HMXBs comprising black holes (hereafter, BHXs) are among the most
powerful X-rays and jet emitters. The potential of BHXs  as an efficient
feedback mechanism at high redshift is supported by observational results from
\citet{Fender2005} which suggest that the
contribution of kinetic energy from BHX jets could be significant compared with
the energy injected by SN, in some cases. \citet{Ramsey2006} studied the
interstellar environment of seven HMXBs, showing that to understand the
ionization and kinematics of the super-shells around these sources, it is
necessary that HMXBs contribute with a significant amount of kinetic energy.
Several works indicate that the energy deposited by BHXs in the form of
kinetic energy could be as high as their bolometric X-ray luminosities
\citep{Gallo2005,Pakull2010,Feng2011,Soria2014}.  %\citep[see][and references therein]{Feng2011}.
Based on these observational evidences, \citet{Justham2012}
conclude that the early injection of energy by these sources
could heat up the ambient gas without expelling it from the galaxy, preventing
the early transformation of gas into stars, and changing the properties of the
ISM where SN events will take place. As we mentioned before, the main impact of
this process is expected to occur at high redshift, where low-metallicity
sources prevail.

In this paper, we implement for the first time, a BHX  feedback model within cosmological
hydrodynamical simulation, in a self-consistent way. Our
simulations include a chemical evolution model \citep{Scannapieco2005} and a
physically-motivated SN feedback \citep{Scannapieco2006}, therefore the enrichment of
baryons can be followed as galaxies are assembled. Hence, our simulations allows us to describe the metallicity
dependence of the sources along the Hubble time. This
gives us a further insight into this problem,  by providing the self-consistent
evolution of the gas cooling, the transformation of gas into stars and
the stellar evolution, with the subsequent chemical enrichment of the
baryons. 
We will  consider very low-metallicity stars as BHX progenitors, since they are
expected to be most energetic and abundant. We explore the effect of BHX
feedback on the cosmic star formation history and on the properties of galaxies
within haloes of different masses. Our results are in agreement to
those reported by \citet{Justham2012} where a semi-analytical model is
developed to study the effects of BHX feedback.

This paper is summarized as follows. In Sect.~\ref{cosm_sim} we describe the
numerical simulations and the implementation of BHX feedback. In Sect.~\ref{s_csfr} we explore the effects of this feedback on the cosmic star formation
history, while in Sect.~\ref{eff_gxy} we study its effect on simulated
galaxies. Finally, we present our main conclusions in Sect.~\ref{concl}.

\section{Numerical Methodology}\label{cosm_sim}

In this section, we describe the main
characteristics of the cosmological code, the simulations, and  the model  developed to incorporate the
BHXs self-consistently in the numerical code.

\subsection{The cosmological code}\label{mod_gen}

We use an extended version of the TreePM/SPH code {\small P-GADGET-3} \citep{Springel2005},
which includes a multiphase model for the gas component, metal-dependent
cooling and SN feedback as described by \citet{Scannapieco2005,Scannapieco2006}.

The multiphase scheme allows the coexistence of low- and high-density gas
clouds, diminishing the problems of over-smoothing of standard SPH. The
main difference is that neighbouring particles are selected not only by a
distance criterion, but also by considering their relative entropy. The
decisions are made on a particle-particle basis, which implies no
mass-dependent parameters to be fixed
\citep[see][for details]{Scannapieco2006}. The multiphase model works
coherently with the SN feedback model, which includes Type II and Type Ia SN
events (hereafter, SNII and SNIa, respectively). When SN events are produced,
metals and energy are distributed within the hot and cold gaseous phases
surrounding the star particle representing a single stellar population. 
The cold phase of the star particle is defined by gas particles with $T < 2 T_{*}$ and
$\rho > 0.1 \rho_{*}$, while the rest of the gas determines the hot phase
($\rho_{*} = 7 \times 10^{-26}\,{\rm g\,cm^{-3}}$ and
$T_{*} = 4 \times 10^{4}$~K, see \citealt{Scannapieco2006} for a detailed
discussion on these parameters). The fraction of energy distributed into the
cold phase is defined by the parameter $\epsilon_{\rm c}$. An exploration of
this parameter and  its  effects is given by \citet{Scannapieco2006,Scannapieco2008}. Here, we
adopt $\epsilon_{\rm c} = 0.5$ as it has been done in previous works which used
this SN feedback model \citep{Scannapieco2009,DeRossi2013,Pedrosa2014}.
Note that these cold and hot gaseous phases are defined for the stellar
populations where SN events are estimated to be produced, and for the only
purpose of distributing metals and energy. They have no direct relation with the
SPH integration itself. Hot gas particles thermalize immediately the SN energy
they receive, whereas cold gas particles build up an energy reservoir. The
feedback energy is stored in these reservoirs until gas particles have enough
energy to change their entropy in order to  match that of their surrounding hot
neighbour media. When this occurs, the reservoir energy is pumped into the
internal energy. This SN scheme is able to produce powerful, mass-loaded
galactic winds as has been shown in previous works
\citep[e.g.][]{Scannapieco2008,Scannapieco2009}.

Stars more massive than $8 M_{\odot}$ are considered as SNII progenitors. We estimate the number of SNII by adopting the
Initial Mass Function (IMF) of \citet{Chabrier2003}. For SNIa, we use the
simple model of \citet{Mosconi2001} where the SNIa rate is estimated by
adopting an observationally motivated ratio ($\sim 0.0015$). The lifetimes of
SNIa progenitors are randomly selected within the range [0.1--1]~Gyr. This model,
albeit simple, allows a good representation of the chemical patterns of the
stellar populations \citep{Jimenez2014}. The initial chemical composition of
the gas component is assumed to be primordial ($X_{\rm H} = 0.76$, $X_{\rm He}
= 0.24$). The chemical model follows the enrichment by twelve isotopes:
$^1$H, $^2$He, $^{12}$C, $^{16}$O, $^{24}$Mg, $^{28}$Si, $^{56}$Fe, $^{14}$N,
$^{20}$Ne, $^{32}$S, $^{40}$Ca and $^{62}$Zn.
% adopting the yields of \citet{Thielemann1993}. 
For SNIa we adopt the nucleosynthesis model of \citet{Iwamoto1999} and for SNII that of \citet{Woosley1995}. Chemical elements are distributed in a
similar fashion as the energy, although 80\% of the new elements are dumped
into the cold phase (i.e. $20 \%$ goes into the hot phase). The cooling functions are metal-dependent \citep{SD93}.

\subsubsection{Numerical experiments}
The simulated volumes represent 14~Mpc comoving-side boxes, and are
consistent with the  cosmology $\Lambda$-CDM with $\Omega_{\Lambda}=0.7$, $\Omega_{\rm m}=0.3$,
$\Omega_{\rm b} = 0.04$, $\sigma_{8}=0.9$, and
$H_{0}=100 h\,{\rm~Mpc^{-1}~km\,s^{-1}}$ with $h=0.7$. Although this is not the
current favourite set of cosmological parameters, we use it because it allow us
to compare these simulations with previous ones. The variation of the
cosmological parameters will not affect the conclusions of this work, which is
related to the effects of feedback on the properties of the ISM and the
regulation of star formation within individual galaxies. For this purpose, we
compare runs of the same initial conditions but with different feedback
mechanisms.

Initially, the simulations contain $230^3$ dark matter particles and
$230^3$ gas particles (we will refer to these simulations as S230).
The dark matter and initial gas particle masses are
$\sim 9.1 \times 10^{6} \, M_{\odot}$ and $\sim 1.3 \times 10^{6}\, M_{\odot}$.
The adopted gravitational softening length is $1.24~h^{-1}$~kpc. 
S230 runs were followed down to $z=0$, but we focus the
analysis on $z > 4$.

Higher numerical resolution runs of $2\times 320^3$ particles were also
performed to investigate the dependence on mass resolution
(we will refer to these simulations as S320). The dark
matter and initial gas mass in this case are $\sim 3.1\times 10^{6} \,
M_{\odot}$ and $\sim 4.9\times 10^{5}\, M_{\odot}$, and the gravitational
softening length  is $0.5 h^{-1}$~kpc.
These simulations were run down to $z\sim 7$ due to their high
computational cost. We adopted the same star formation and feedback
parameters used in S230. As a consequence, the star formation
activity starts at higher redshift since smaller systems are better
resolved. As a result, the impact of BHXs occurs also at higher
redshift in S320 than in S230. Nevertheless, this behaviour does not affect our
analysis since we are always comparing runs with and without BHX
feedback with the same resolution to draw conclusions.
%S320 experiments were performed for the only purpose of
%analysing the effects of resolution on the BHX modelling. 

We have run the same initial condition with SN feedback (S230-SN and
S320-SN) and with SN+BHX feedback (S230-BHX and S320-BHX) in order to assess
the effects clearly. 
Although in this work, we only  discuss  the most successful
implementation, we have run the same initial conditions varying the parameters
of the BHX models, as explained in the following section.

\subsection{The stellar feedback model for BH-HMXBs}\label{feed}

We are interested in studying the effects of the kinematic feedback of BHXs
 originated from very low-metallicity progenitors. As these metal-poor
 progenitors  are  expected to form  principally  in the Early
Universe, their contribution should be relevant for the regulation of the star
formation in the early stages of galaxy evolution.  As mentioned in the
Introduction, we considered previous results from 
\citet{Power2009} and \citet{Justham2012} to develop our
model, which is fully implemented within our SPH code.

We will consider BHX populations rather than individual sources due to 
the numerical resolution of our simulations.
We assume that massive stars 
with  metallicities in the range $Z_{\rm EMP} = [0,10^{-4}$) will give origin to \textit{extremely metal-poor} 
progenitors of BHXs (see below for more details). This metallicity range is taken as indicative, since there is no clear
theoretical or observational value to be used. Nevertheless, theoretical works predict larger formation
rates for $Z < 0.01 Z_\odot$ \citep[e.g.][]{Linden2010}.
The upper limit
represents the characteristic value of the oldest Population II stars \citep{Belczynski2004a}.
We would like to stress that although the upper metallicity limit was selected with this criterion,
sensitive variations should not affect strongly our results.
Note also that we are not modelling Pop~III stars. We assume our sources to be very low metallicity stars but
not representative of the first stars. Pop~III sources are beyond the scope of
this paper and  would demand the treatment of other physical mechanisms
which are not included in our code \citep[e.g.][]{Chen2014}. Our
model includes the $Z = 0$ stellar populations since there are not efficient
mixing processes within the SPH kernel. Their treatment  
should increase the metallicity floor quickly from  $Z = 0$. 

 In order to estimate the number of BHs produced in each young stellar population, we adopt
the model of \citet{Georgy2009} for the evolution of massive stars. This model includes stellar rotation
and a dependence of the type of compact remnant on the metallicity of
the progenitor. The fraction of massive stars that end their life as BHs increases with decreasing metallicity.
Therefore, for each young stellar population we estimate the number of
BHs produced by adopting the IMF of \citet{Chabrier2003}. We consider 
stellar masses in the range of $30-120 M_{\odot}$  and metallicities
consistent with the defined  $Z_{\rm EMP}$ interval.
We assume that $\sim 20\%$ of low-metallicity BHs will end up in binary systems, 
producing a BHX ($f^{\rm EMP}_{\rm BHX}=0.20$). This value is roughly
comparable to  the $30~\% $ reported
by \citet{Power2009}, after correcting by the different IMF
adopted.

As we mentioned before, in BHXs the BH accretes mass from its companion star releasing
a significant amount of energy in the form of radiation and outflows.
Observational evidence from shock-ionized bubbles in the environment 
of these sources ($\sim$~50--300~pc) suggests that their mechanical output energy 
would be around $\sim10^{52}\,{\rm erg}$ \citep{Soria2014}.
On the other hand, many authors have reported that the kinetical power from accreting black holes 
could be as high as their Eddington luminosity limit \citep{Pakull2003,Pakull2010}.
For example, for Cygnus~X-1 \citet{Gallo2005} have found that its jet   
transforms its total power into kinetic energy, and that the latter could be greater than its 
bolometric X-ray luminosity.
Similar results have been found by \citet{Soria2010} for the microquasar S26 in NGC~7793, and
by \citet{Soria2014} for an accreting BH in M83.

Motivated by these observations,  we assume that the energy deposited by low-metallicity
BHXs in the form of kinetic energy is $E_{\rm BHX} = 10^{52}\,{\rm erg}$ \citep{Mirabel2011,Feng2011}.
Assuming  that a similar amount of energy is radiated in the X-ray band
during $\sim 3\,{\rm Myr}$, the mean X-ray luminosity per source is
$L_{\rm X, source} \sim 10^{38}\,{\rm erg \, s}^{-1}$, which is typical of HMXBs.
We also tested larger and smaller energy  values. Energies of the order of
$E_{\rm BHX} = 10^{53}\,{\rm erg}$ sweep the gas in haloes too efficiently while
lower values ($E_{\rm  BHX} \sim 10^{51}\,{\rm erg}$) produce no significant impact on the
regulation of the star formation activity.

Hence, for each young stellar population with  metallicity in the range $Z_{\rm EMP}=[0,10^{-4})$, 
we estimate the number of BHXs as  $N_{\rm BHX}=f^{\rm EMP}_{\rm BHX}\times N_{\rm BH}(Z)$
where $N_{\rm BH}(Z)$ is the number of BHs produced according to \citet{Georgy2009} and $f^{\rm EMP}_{\rm BHX}=0.20$,
 as we discussed before.
Each of these $N_{\rm BHX}$ sources releases $E_{\rm BHX} =
10^{52}\,{\rm erg}$ into the ISM. We assume that this energy will be
efficiently thermalized by the surrounding gas clouds where the
the events took place.  This hypothesis is supported by observational evidences that show that these events are expected to  affect the region
surrounding the stellar progenitors \citep[$\sim 300$~pc,
e.g.][]{Pakull2010}. As a consequence, the released energy is dumped into the
internal energy of the nearby cold gas phase, contributing to build up their
energy reservoir.

\section{The cosmic star formation rate density} \label{s_csfr}

\begin{figure}
\centering
\includegraphics[width=0.5\textwidth]{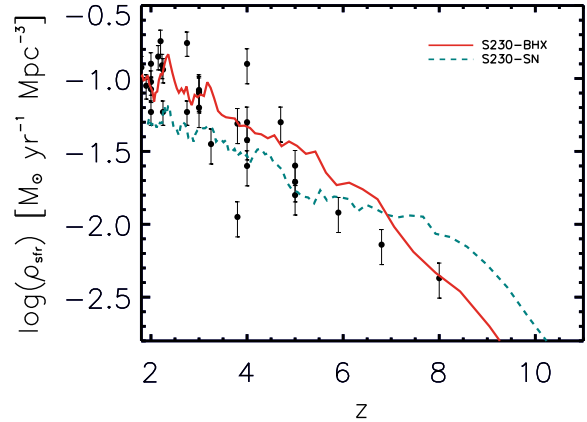}
\caption{The cosmic star formation rate density estimated from simulations with
SN feedback (S230-SN, blue dashed line) and  with both SN and BHX
feedbacks (S230-BHX, red solid line). For comparison, the observations compiled by \citet{Behroozi2013}
are also included (black filled circles). The two runs share the same
cosmology, and the SF and SN feedback parameters. The cosmic SFR is diminished by
BHX feedback at early epochs ($z \gtrsim 7$) through cold gas heating, and
boosted at lower redshift as more gas is  available to feed the  star
formation activity.}
\label{cSFR}
\end{figure}

The SN and  BHX feedback mechanisms modify the thermodynamical properties
of the ISM, thus affecting the transformation of gas into
stars, which is the main effect we analyse in this paper.
 In order to have a global picture of their impact, we estimate the cosmic star formation rate density
(cSFR) for both simulations, S230-SN and S230-BHX. As  shown in Fig.~\ref{cSFR},
the confrontation of the cSFR with observational results compiled by
\citet{Behroozi2013} suggests that the treatment of BHX feedback worked in the
expected way, yielding an improved description of the observations.
The factor $f^{\rm EMP}_{\rm BHX} \sim 0.20$ was actually chosen to be able to represent this
observational relation\footnote{For the adopted $E_{\rm BHX} =
10^{52}\,{\rm erg}$, we tested $f^{\rm EMP}_{\rm BHX}$ between $\sim
1\%$ and $\sim 20 \%$, finding  $f^{\rm EMP}_{\rm BHX} \sim 0.20$
reproduced the observed trend (Fig.~\ref{cSFR}).}. However, this factor might vary if the $E_{\rm BHX}$ is
changed accordingly so that the total amount of released energy per
population remains within the same values. Otherwise, the effects on
the ISM are found to be too weak or too strong. The fact that it agrees roughly with
that derived by \citet{Power2009} is encouraging.

\begin{figure}
\centering
\includegraphics[width=0.5\textwidth]{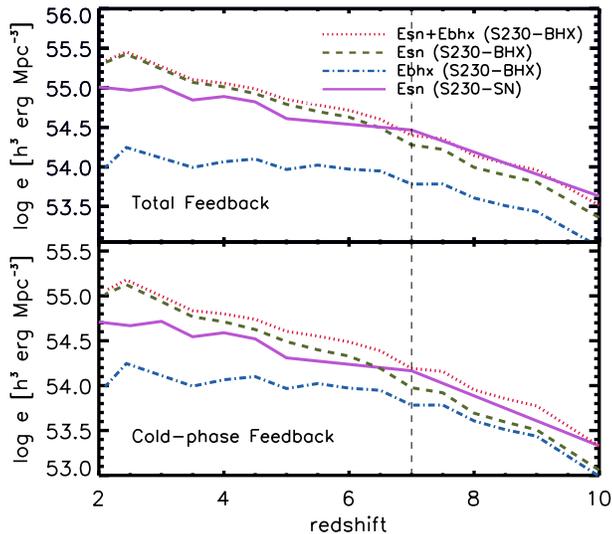}
\caption{{\it Upper panel:} Comoving energy density released by different feedback mechanisms
as a function of redshift: SNII in S230-SN (magenta solid line), SNII in
SN230-BHX (green dashed line), and BHXs in S230-BHX (blue dot-dashed line).
The total comoving energy density released by all agents in S230-BHXs (BHXs~+~SNe) is also plotted
(red dotted line). BHXs release a significant amount of energy, of $\sim$30\%
of the SNII+BHX feedback at $z\gtrsim7$, but its fractional contribution decays for
lower redshifts, down to $\sim$10\% at $z\sim4$. {\it Lower panel:} Same as
above, but taking into account only the energy injected into the cold gas phase
of the ISM. The contribution of BHXs is larger in this case because they dump
all their energy into the cold phase.}
 \label{energy2b}
\end{figure}

In order to understand the combined effects of the two feedback mechanisms, in Fig.~\ref{energy2b} (upper panel),
 we show the  comoving energy density ($e$) released
by SNII and BHXs, as a function of redshift, 
for both S230-SN and S230-BHX. We only  considered SNII contributions  since they
also originate from massive stars. SNIa events will also contribute but with
a larger time-scale and lower rate. 
%For simplicity they are not
%included in these estimations.

Both the SNII and the BHX energy contributions grow with redshift, due to
the increasing cosmic star formation activity. However, BHX energy feedback
grows at a slower rate than SNII one, due to the metallicity
dependence of the progenitors. As the metallicity of
the ISM increases, the number of BHX events diminishes, making
 SNe the dominating feedback mechanism. 
As can be seen from Fig.~\ref{energy2b}, BHX feedback is
significant for $z > 7$. As an example, at $z \gtrsim 7$ it represents  $\sim 30\%$ of the total
energy (SNII+BHX) released into the ISM. For lower redshift, the contribution of BHX
feedback progressively decays, reaching less than $\sim 10\%$ of the
total at $z \sim  4$ ($\sim5\% $ at $z \sim 2$).
The total $e$ released by SNII and BHX together is comparable to that of S230-SN for $z
\gtrsim 7$, while for  lower
redshift, that of S230-BXH is greater due to the SFR enhancement
triggered in this simulation, as can be  seen in
Fig.~\ref{cSFR}\footnote{The impact of BHX feedback on the global SF
  activity when the numerical resolution is increased is similar,
  although as explained before, S320 runs formed more stars for the
  same set of SN and SF parameters (see  Section 2.1.1). By comparing the energy
released by BHX
and SNII events in S320-SN and S320-BHX, we found a maximum  of $\sim
30\%$ contribution from BHX feedback, that  decays
rapidly with redshift as the ISMs are chemically enriched,  in a
similar fashion as
reported for  S230 in Section 3. The specific redshift at which BHX feedback is maximum 
over SNII feedback is a statement which  depends on the box-size.
Our simulated box correspond to a small field region, so a larger volume will be 
required to make a full statement on this aspect.}. We will come back to this point later on.

The  SNII and BHX energy release should affect the regulation of
the star formation in different ways, depending on the energy available to heat
up the cold gas clumps where stars are born. Fig.~\ref{energy2b} (lower panel) displays the
energy released by SNII and BHXs to the cold gas phase, as a function of
redshift. For S230-BHX, the amount of energy dumped by BXHs at high redshift
is similar to that of SNII. The total amount of feedback into the cold gas is
then larger in S230-BHX than in S230-SN, producing an early heating of the gas
that diminishes the star formation activity. In fact, this can be appreciated
from the cSFR, shown in Fig.~\ref{cSFR}
for both simulations. As expected, in S230-BHX the BHX feedback diminishes the
star formation activity for $z \gtrsim 7$ with respect to that of S230-SN,
because it provides a larger amount of energy used directly in cold gas
heating. The lower star formation rate in S230-BHX also explains the lower SNII
feedback in this simulation (Fig.~\ref{energy2b}). We note that the star
formation activity in S230-BHX increases for $z < 7$ faster than in S230-SN,
reaching values comparable to observations \citep{Behroozi2013}. Hence,
the impact of BHXs on the regulation of the star formation is boosted to lower
redshift by making gas available to star formation at late times.

\section{The effects of BHX feedback on simulated galaxies} \label{eff_gxy}

In order to assess the effects of BHX feedback on the regulation of the
star formation in galaxies of different masses, we first constructed 
galaxy catalogues from S230-SN and S230-BHX.
We use the Friends-of-Friends technique to select the virialized structures, and
the {\sc subfind} algorithm \citep{Springel2001} to identify the
substructures within
the virial radii. We only consider simulated galaxies resolved with
more than 500 particles. In order to test the robustness of our results against
numerical resolution, we  run a higher numerical resolution initial condition increasing the mass resolution by a factor of eight. We acknowledge the fact
that our simulated volume is small to provide a complete
description of the galaxy populations. However, by comparing the same
initial condition run with different feedback agents, we are able to underpin the effects of these agents within the mass range of $\sim 10^9 - 10^{11}\, M_{\odot}$. Hence, the analysis should
be taken on  individual galaxy basis, and as a first step toward
understanding the impact of BHXs as a possible feedback mechanism.

We carry out an analysis of the simulated galaxies from $z\sim 9$ to
$z\sim 4$. For each available snapshot of the simulations,  simulated galaxies are defined at the optical radius which corresponds
to that enclosing $83 \%$ of the baryonic mass of the selected
system. The dark matter halo masses are calculated at
the virial radius.

% ***************************************************
\subsection{The stellar-to-dark mass ratio}\label{msmh_res}

\begin{figure}
\centering
\includegraphics[width=0.4\textwidth]{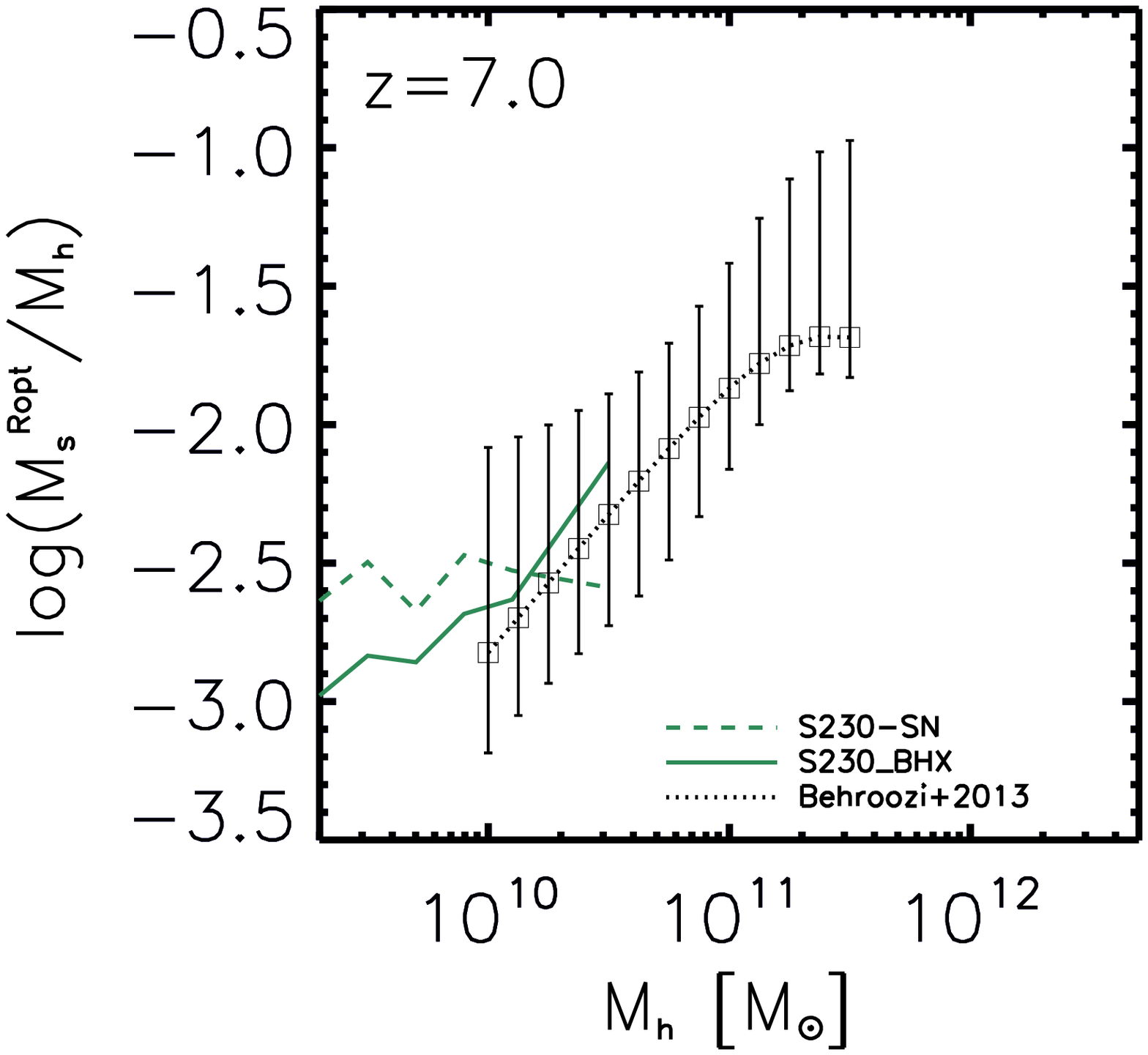}
\includegraphics[width=0.4\textwidth]{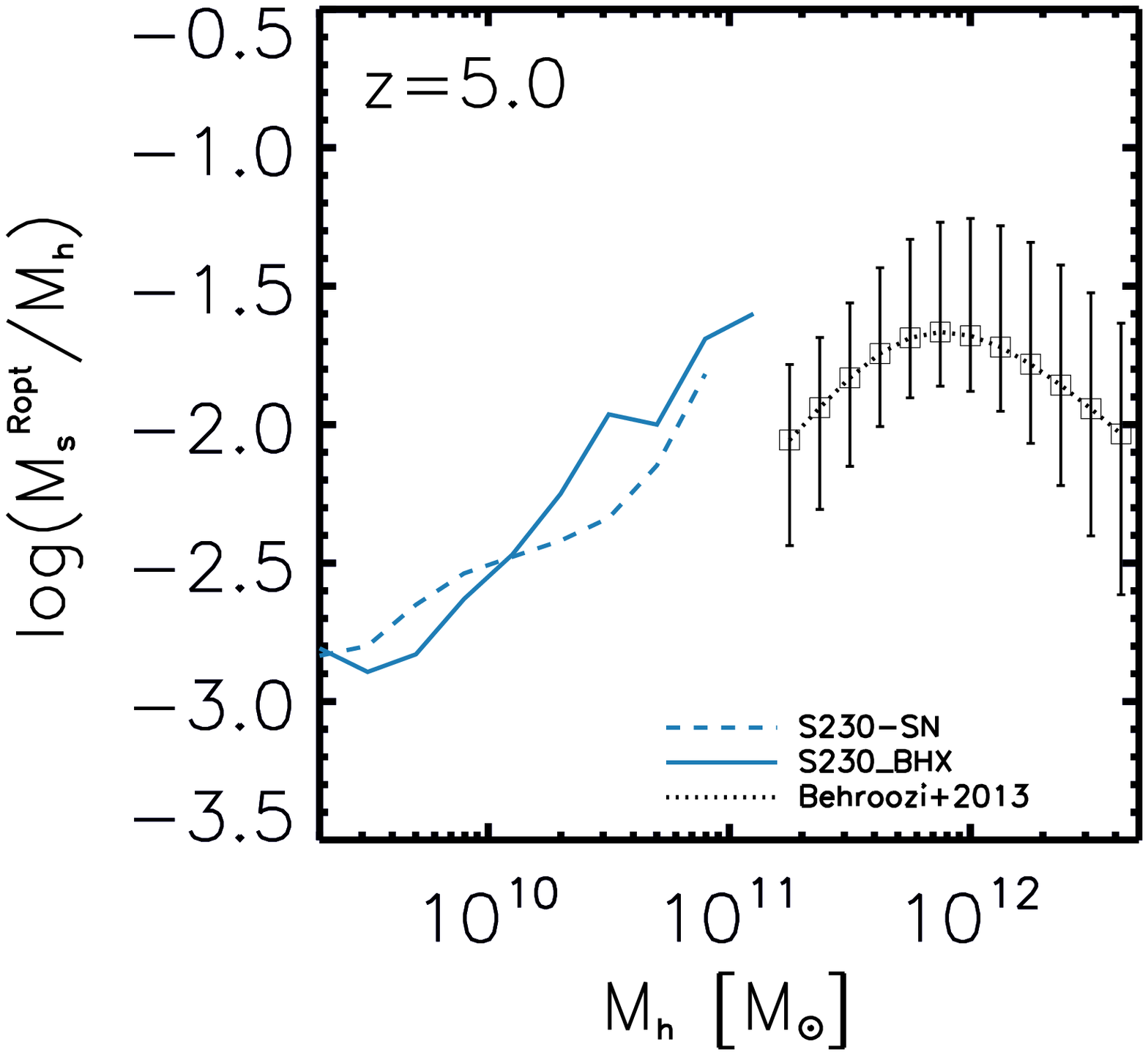}
\includegraphics[width=0.4\textwidth]{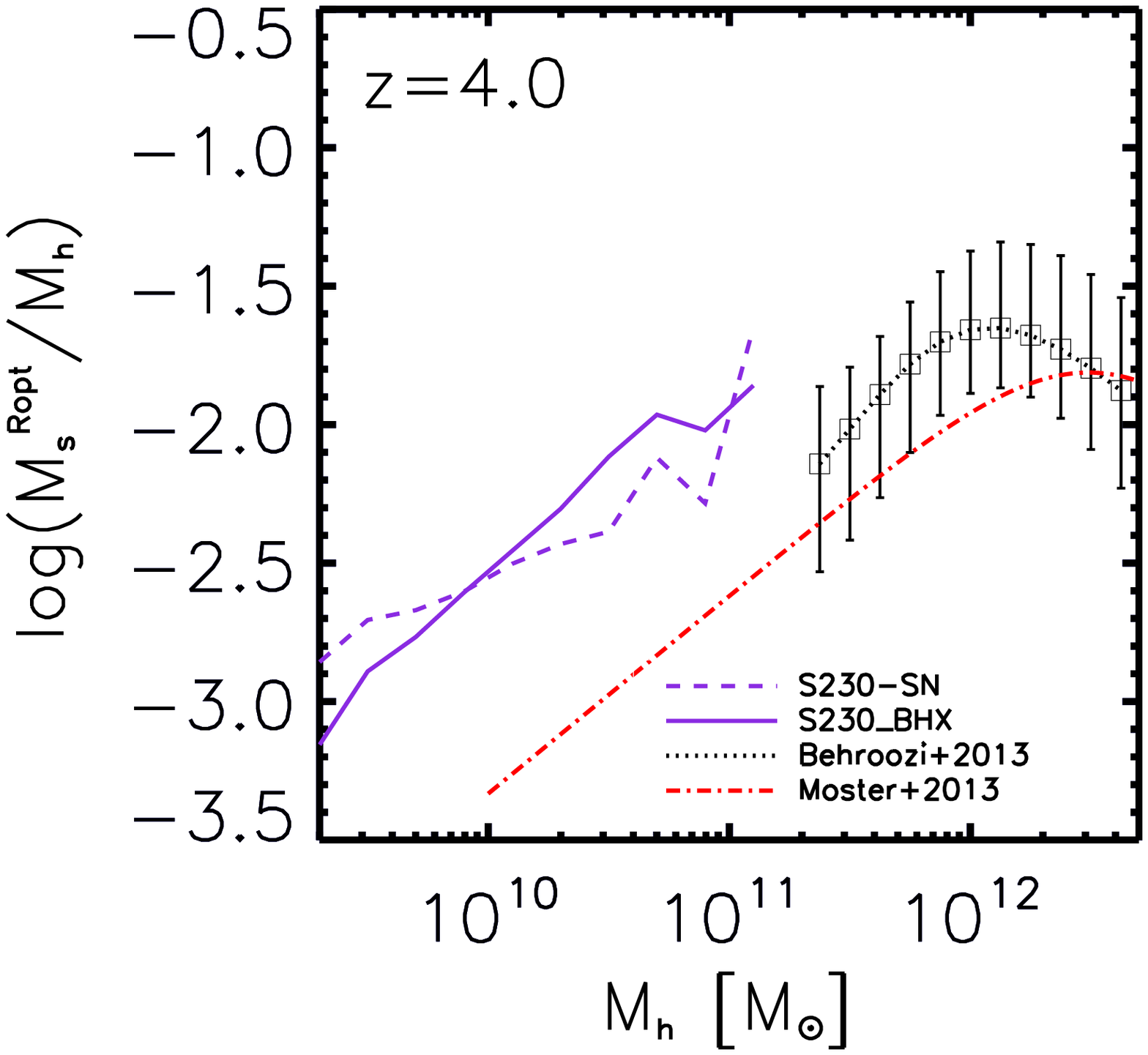}
\caption{Mean stellar-to-halo mass ratio as a function of halo mass,
for redshifts $z \sim 7 $, 5 and 4. We estimate the stellar mass of a galaxy as
the total mass of stellar particles inside the optical radius. Dashed lines
represent galaxies in S230-SN, while solid lines show those in S230-BHX. 
The results of abundance matching techniques taken from \citet{Behroozi2013} are
plotted in black as open squares plus a dotted line. For $z \sim 4$ we include
also the extrapolation of the data of \citet{Moster2013}
(dash-dotted red line).}
\label{mh_ms}
\end{figure}

First, we study the ratio between the mean stellar mass and  their virial
halo mass of the simulated galaxies as a function of the latter. The estimated relations for S230-SN and
S230-BHX are shown in Fig.~\ref{mh_ms}. For comparison, we also show the
results obtained using the abundance matching technique by \citet{Behroozi2013}
at redshift $z \sim  7,  5, 4 $, and by \citet{Moster2013} at redshift
$z \sim  4$. In the case of \citet{Moster2013} data, we extrapolate their
stellar-to-halo mass relation to lower halo masses.
We note that the simulations tend to resolve smaller virial haloes compared to
those shown by \citet{Behroozi2013}, except for $z \sim 7$ where there is a
better match of the mass range. Therefore, we extrapolate the trends to draw
conclusions.

At $z \sim 7$, there is an excess of stars in small haloes in S230-SN
compared to those detected in S230-BHX. This excess originates in the very efficient
transformation of gas into stars at the early stages of evolution in
S230-SN as also reported in 
previous works \citep[e.g.][]{Sawala2011,DeRossi2013}. Although  our SN
feedback model is successful at producing galactic outflows and fountains,
which regulate the star formation, it is less efficient in the first
stages of galaxy formation.
When the energy released by BHXs is added to the SN feedback,
  then a decrease  in the star formation activity at high redshift is detected. In Fig.~\ref{energy2b}, we
compared the relative importance of both feedback mechanisms as a function of
redshift. In Fig.~\ref{mh_ms} we can appreciate the effects on
individual galaxies. In fact, the regulation of the star
formation by BHX events is more efficient in low mass haloes, steepening the
relation between the stellar-to-halo mass ratio and the halo mass of the
galaxies. In haloes less massive than $\sim 10^{10}\, M_\odot$, simulated
galaxies in S230-BHX have less stellar mass than those in S230-SN. For larger
halo masses instead, the simulated galaxies in S230-BHX reach larger fractions
of stars compared to the SN feedback run. This trend is produced because
simulated galaxies in these haloes have larger amount of gas to feed
the subsequent star
formation process. This can be explained as a consequence of the impact of
BHXs at early times, when the systems were smaller and could be affected by BHX
feedback.

We note that although the cSFR of S230-BHX reproduces the observations compiled
by \citet{Behroozi2013}, there seems to be still more stars per dark
matter halo at $z \lesssim 7$. However, the slope of stellar-to-halo mass
relation agrees better to the extrapolation of observations for  lower masses.
This is also valid when compared to \citet{Moster2013} at $z \sim 4$.

\subsection{Gas fractions} \label{gas_res}

\begin{figure}
\centering
\includegraphics[width=0.5\textwidth]{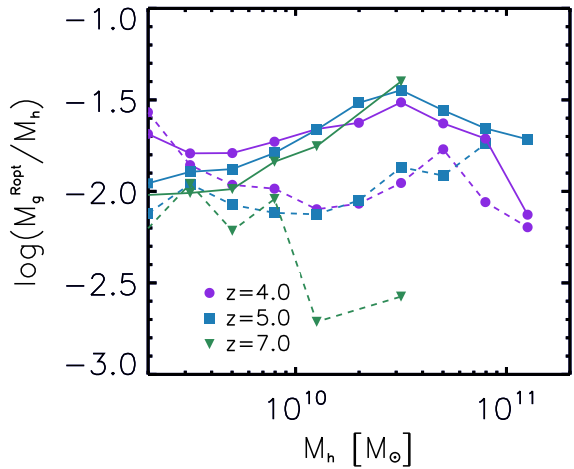}
\caption{Gas-to-halo mass ratio as a function of halo mass for simulated
galaxies in S230-BHX (solid lines) and S230-SN (dashed lines). The relations
are shown for three different redshifts: $z = 7$ (green triangles), 5 (blue
squares), and 4 (violet circles). Gas mass is estimated as the total mass of gas particles enclosed
within the optical radius of a galaxy at a given redshift.}
\label{mg_mh}
\end{figure}

The impact of the SN and BHX feedback on the stellar mass of galaxies can be also
studied  by comparing the ratio between the mean gas mass and the virial halo mass.
 Fig.~\ref{mg_mh} shows this relation as a function of the halo mass
for galaxies in S230-BHX and S230-SN, and for the same redshift range
displayed in 
Fig.~\ref{mh_ms}. The gas mass of the simulated galaxies is estimated as the
total mass of gas particles inside their optical radii.

As can be seen from Fig.~\ref{mg_mh}, simulated galaxies in S230-BHX have higher
gas masses than those in the simulation without BHX feedback, regardless of
their host haloes. At $z \sim 7$, we detect the smaller gas fraction in S230-SN
for $M_{\rm h} > 10^{10}\, M_\odot$. This result seems to be at odds with the
trends shown in Fig.~\ref{mh_ms}, under the hypothesis of a close-box
model. Indeed,  if
there were no gas infall or outflow, then the galaxies with the larger stellar
mass fraction should have had the smaller gas fraction. However, this is not what
we see  in Fig.~\ref{mg_mh}. 
%while for lower redshifts, there are similar gas fractions. 
The fact that galaxies in S230-BHX have larger fraction of gas within their
optical radii suggests that their counterparts in S230-SN have lost larger fractions of
their gas reservoirs and hence, the subsequent star formation is lower. If
this were the case, then galactic outflows should have carried out more material
to the circumgalactic medium (CGM) of these galaxies, increasing its
temperature.

\begin{figure}
\centering
\includegraphics[width=0.5\textwidth]{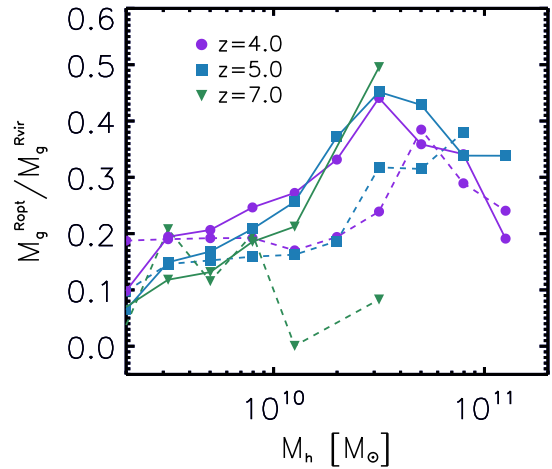}
\caption{Fraction of the total gas within the virial radius retained
  within the optical radius, for simulated galaxies in S230-BHX (solid
  lines) and in S230-SN (dashed lines) at $z = 7$ (green triangles), 5 (blue
squares), and 4 (violet circles).}
\label{mg_mh2}
\end{figure}

In order to analyse this feedback loop between the ISM and the CGM of
the simulated galaxies, we estimate the fraction of the total gas mass
within the virial radius which remains within the optical radius in
both simulations. As can be
seen in Fig.~\ref{mg_mh2}, in S230-BHX,  simulated galaxies in haloes
with  virial masses in the range $10^{10} \, M_\odot \lesssim M_{\rm h} \lesssim 10^{11} \,
M_\odot$ are able to retain larger amounts
of gas than their counterparts in S230-SN. For smaller and larger  haloes,
both simulations show the same level of remnant gas within the
optical radius. In the quoted
halo-mass range, galaxies in S230-SN have been able to loose a
larger fraction of gas, which can be understood considering their higher
early star-formation activity. As a consequence, gas outflows are
triggered, expelling larger amount of gas.
This produces a decrease of the star formation activity afterwards
since there is less gas to fuel it. In fact, as shown in Fig.~\ref{cSFR}, the star
formation activity of S230-SN at low redshift remains lower than that estimated
by \citet{Behroozi2013}, whereas it is higher at high redshift. Hence, this suggests that it will 
not be  possible to reach an agreement if only SN feedback is present \citep[see also][]{Stinson2012}. Increasing
the SN feedback to decrease the star formation at high redshift would make
the discrepancy at lower redshift even larger. It is the action of the
BHX feedback that helps to regulate the star formation at very high
redshifts, which then results in the desired trend at lower ones.

%;;;;What is this velocity threshold

As we can see from Fig.~\ref{mh_ms}, there seems to be a halo mass threshold
below which BHXs are efficient at  modulating the
star formation.  This is approximately $M_{\rm h} \sim 10^{10}\, M_\odot$,
which corresponds to circular velocities $V_{\rm vir} \sim 40\, {\rm km \, s}^{-1}$ or a 
virial temperature of $T_{\rm vir} \sim 50,000\, {\rm K}$ (assuming primordial
abundances). For galaxies below this threshold,
the BHX feedback is strong enough to heat the ISM to $T \sim T_{\rm vir} $, decreasing the
star formation activity without expelling a significant fraction of  gas mass.

\begin{figure*}
\centering
\includegraphics[width=0.45\textwidth]{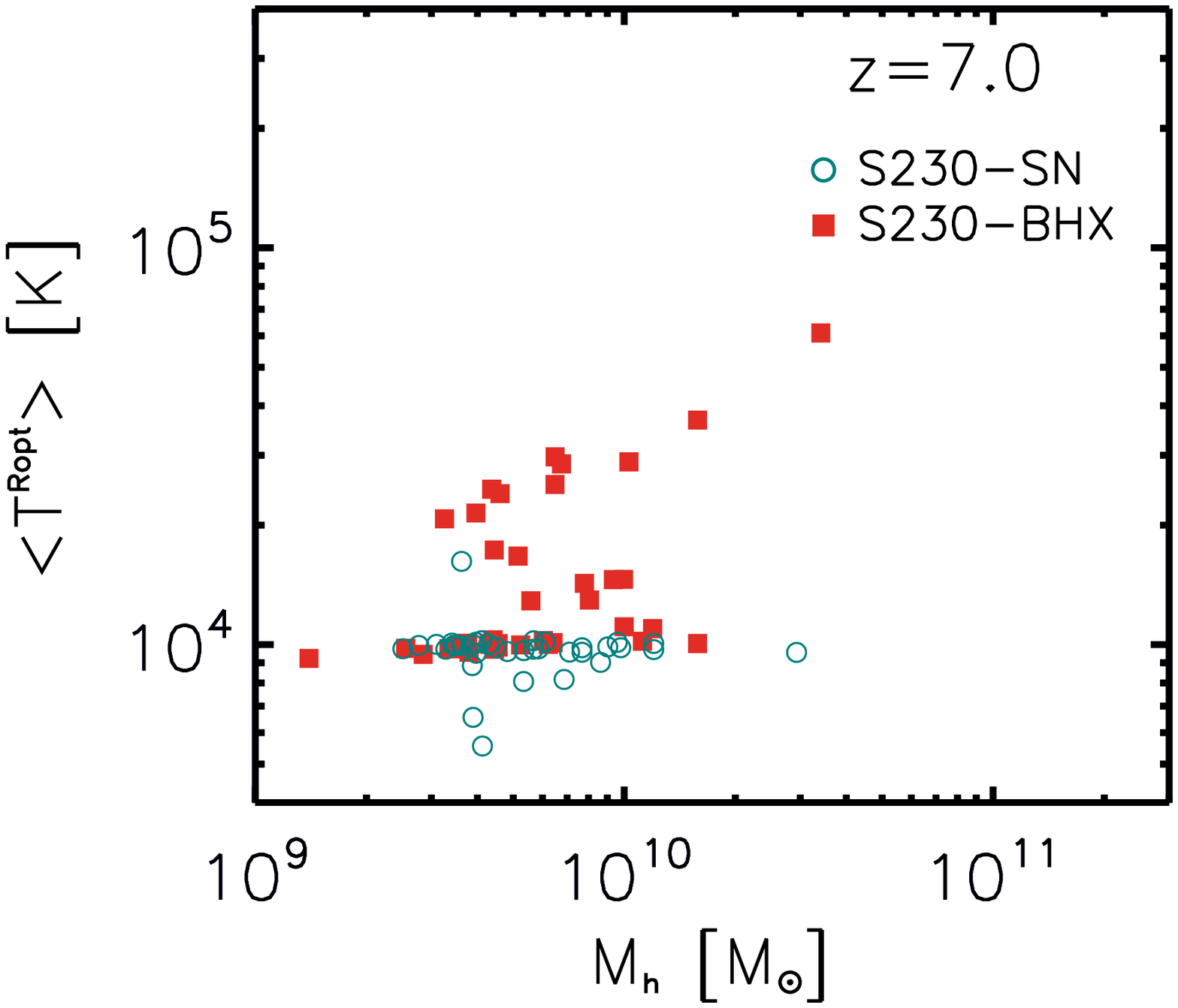}
\includegraphics[width=0.45\textwidth]{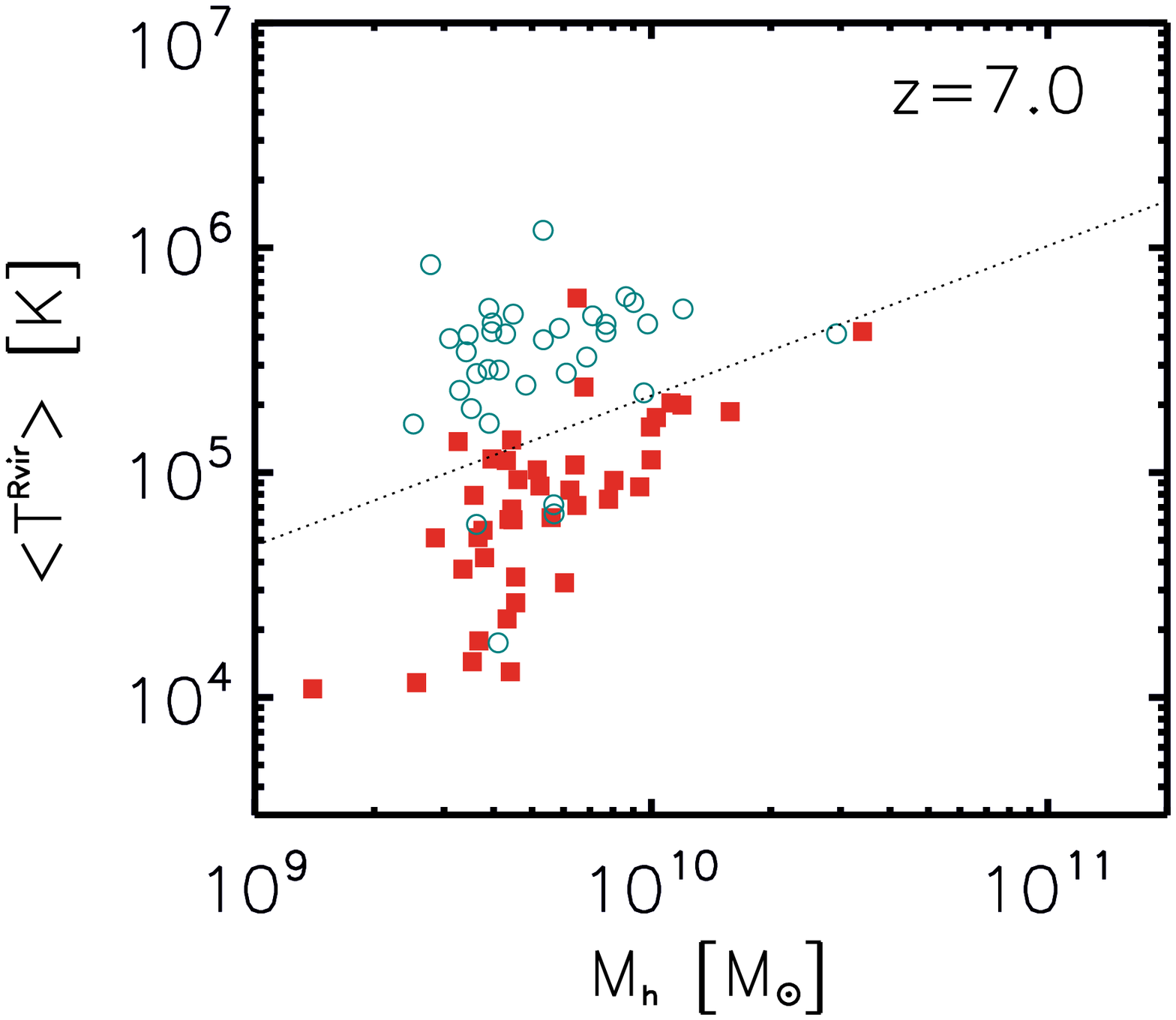}
\includegraphics[width=0.45\textwidth]{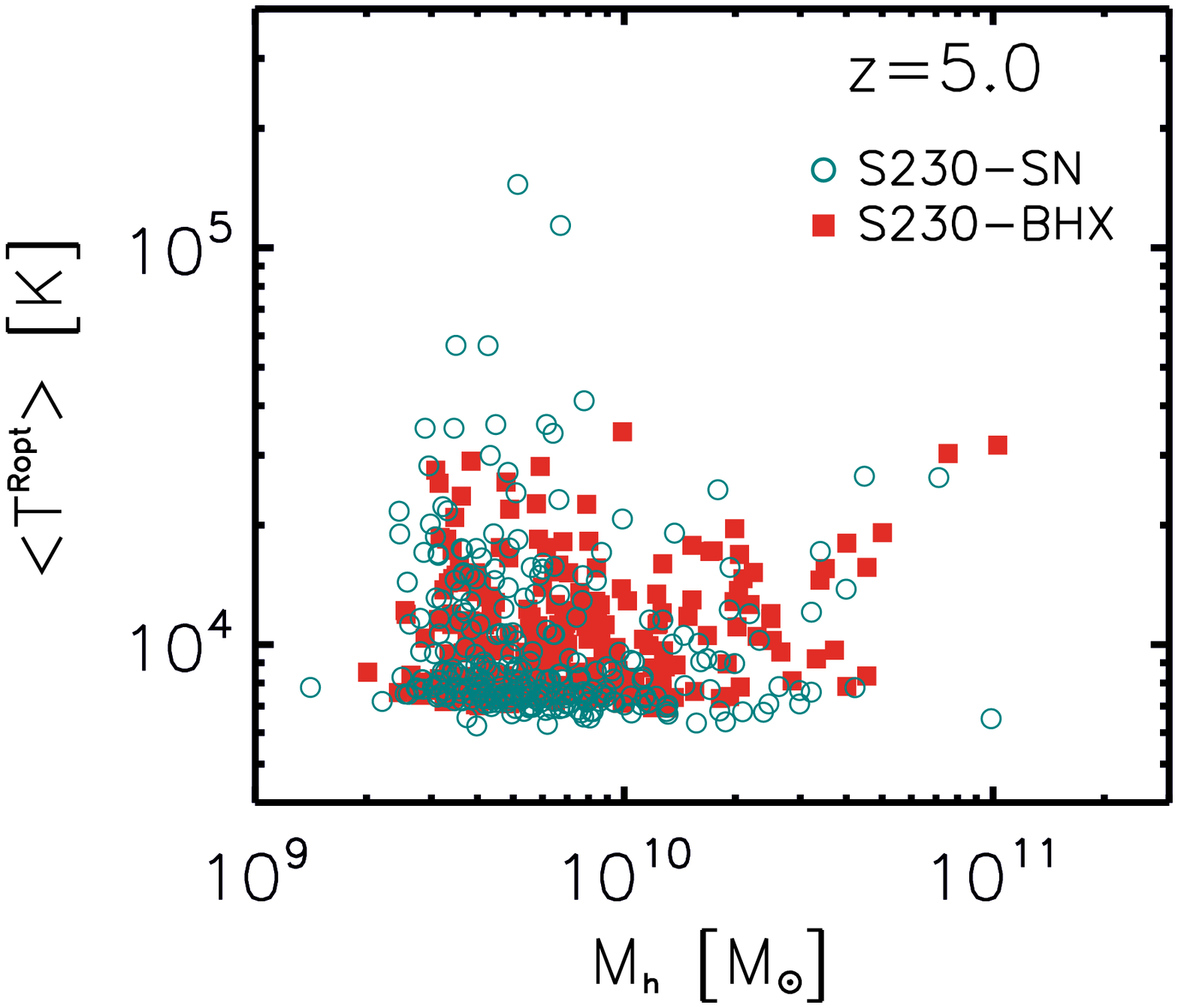}
\includegraphics[width=0.45\textwidth]{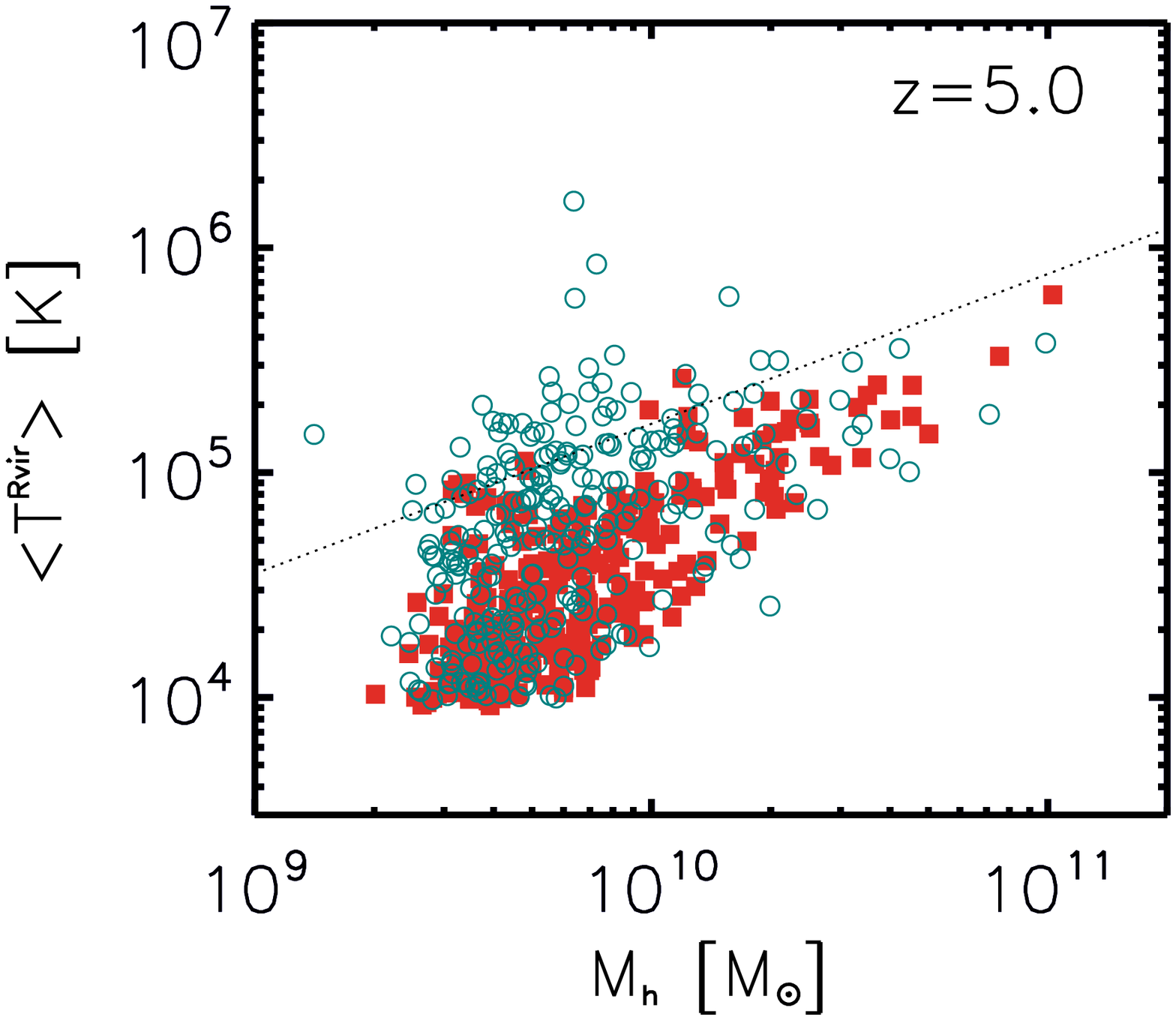}
\caption{Median temperatures of the gas component within the optical
radius (left panels) and within the virial radius (right panels), for
galaxies in S230-BHX and S230-SN at $z = 7$ and $z =  5$.
The dotted lines represent the relation between virial mass and virial temperature.
Note that at $z = 7$, simulated galaxies in S230-BHX have 
hotter interstellar media than their counterparts in
S230-SN. Conversely, the circumgalactic media of galaxies in S230-SN
get  hotter sooner, as a consequence of the hot material transported by
outflows. These outflows are triggered by the more violent SN feedback,
produced by the larger SFR taken place in this run at very high redshift.}
\label{temper}
\end{figure*}

To illustrate this, in Fig.~\ref{temper} we show the mean temperature of the gas within the optical
(left panels) and the virial (right panels) radius, for simulated galaxies in
both runs, at $z \sim 7$ and $z \sim 5$. The run with BHXs shows  hotter
environment within the optical radius, although with temperatures smaller
than $T_{\rm vir}$. Conversely, galaxies in S230-SN have lost 
gas by SN feedback. In fact, SN feedback has been successful at
building a hot CGM even at these high redshifts.
The CGMs are  colder in the run with S230-BHX, since the total
stellar mass formed per halo is not enough to drive powerful winds yet.
The BHX feedback has contributed to heat up the gas within the galaxies,
decreasing the mass of
new-born stars in haloes with $M_{\rm h} < 10^{10} \, M_\odot$. Therefore, as they
grow by hierarchical clustering, they will be able to have  larger gas
reservoirs to continue their star formation activity. 
Later on, as the SN rate increases, the SN feedback contributes to heat
up the CGM of galaxies in S230-BHX.

\subsection{The stellar mass of galaxies}
\label{stellarmass}

\begin{figure*}
\centering
\includegraphics[width=0.4\textwidth]{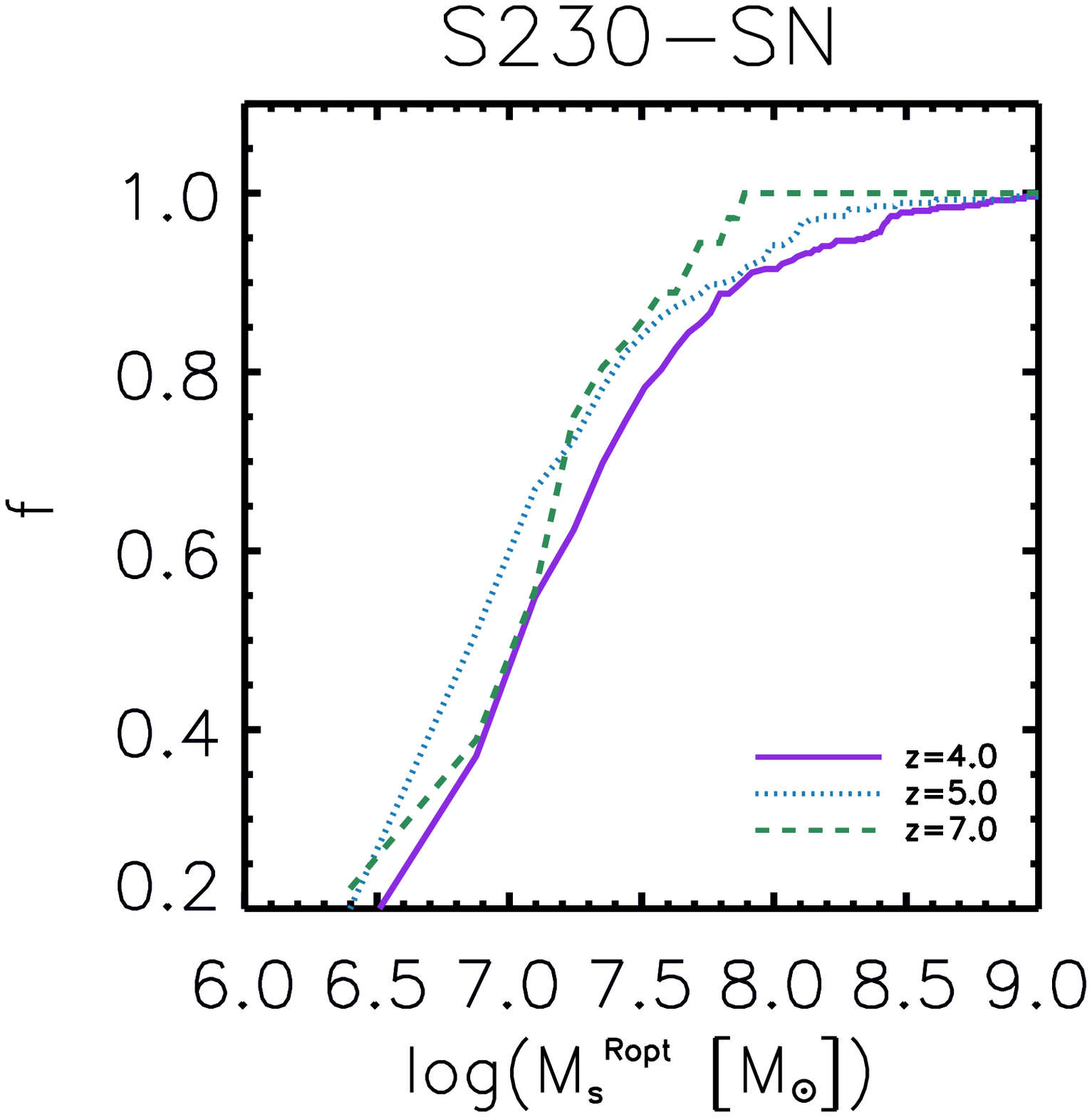}
\includegraphics[width=0.4\textwidth]{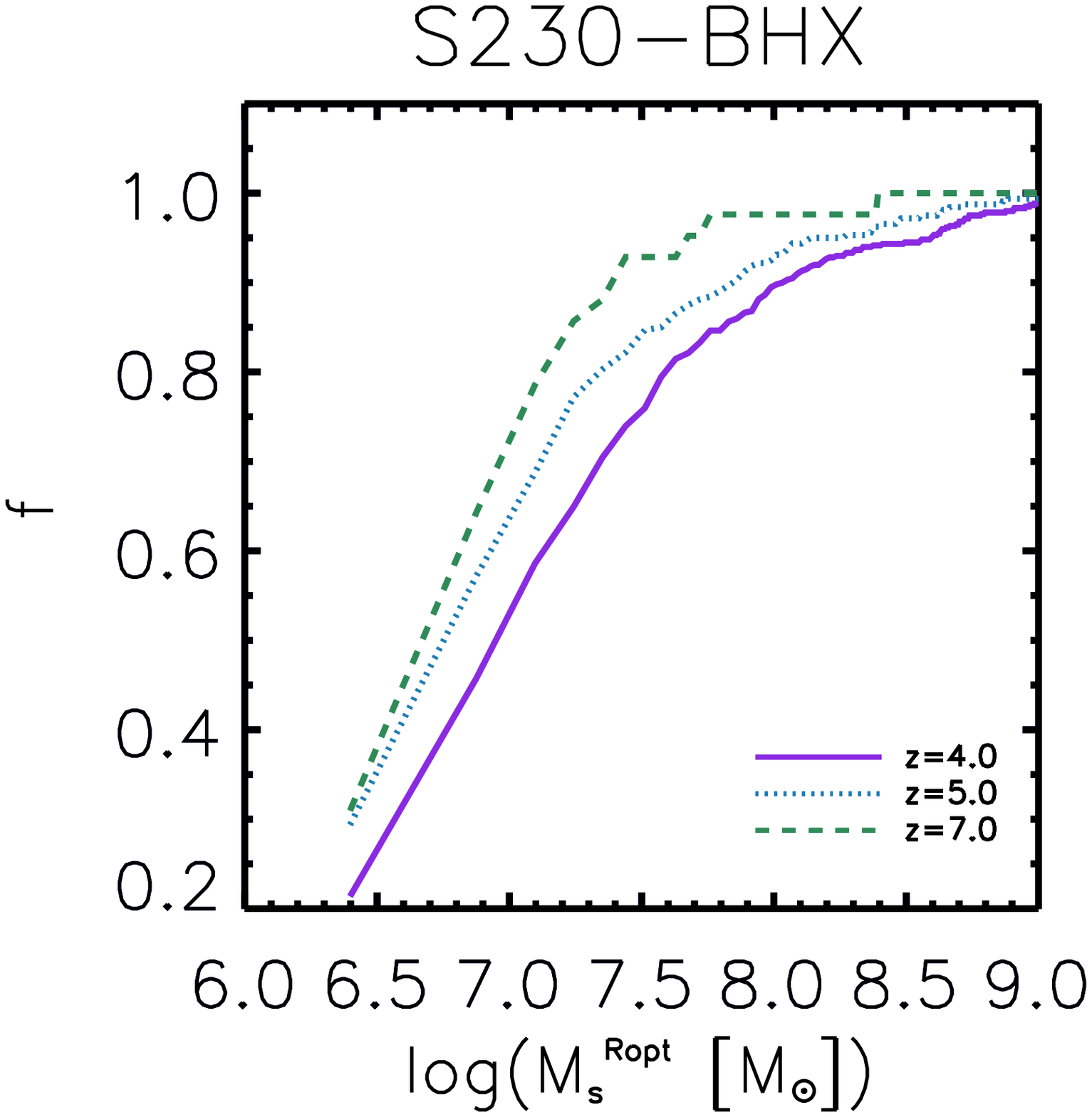}
\caption{Cumulative distribution of the stellar mass within the optical
  radius at $z\sim 7$  (green dashed line),  $z\sim 5$  (blue dotted line) and  $z \sim 4$  (violet solid line).
\textit{Left:} Simulation without BHX feedback. \textit{Right:} Simulation with BHX feedback.}
\label{histogramas}
\end{figure*}

To assess how the effects of BHX  feedback modifies the
stellar mass of galaxies, we   estimate
the  cumulative number of galaxies as a function of their stellar
mass.
As it can be seen from  Fig.~\ref{histogramas}, the simulations behave
differently. The simulation with BHX feedback shows a notable evolution 
with redshift, with $\sim 80\% $ of the galaxies exhibiting  stellar masses
smaller than $10^7 \, M_\odot$ at $z \sim 7$, percentage which decreases
to  $\sim 50\%$ at $z \sim 4$.
 Conversely, the simulation with only SN feedback does not present
a clear evolution in this redshift range. On average, $\sim 50-60\%$ of the
galaxies have stellar masses smaller than $\sim 10^7 \, M_\odot$.
These trends confirm that galaxies in S230-BHX form less stars in
small galaxies than those in S230-SN.
Hence, BHX thermal feedback affects directly small galaxies but indirectly
larger ones.
Those mostly affected have stellar masses between $10^7$ and $10^8$
 $M_\odot$ and inhabit haloes with $V_{\rm vir} \sim 40$ km~s$^{-1}$.

\subsection{Testing numerical resolution}
\label{s320_res}

In order to test the robustness of our trends against numerical resolution,
we analyse the simulated galaxies in S320-SN and S320-BHX at $z\sim 7.5$ and $z\sim 8$.
Galaxies have been selected and analysed
by  using the same criteria, although the minimum number of particles was
increased to 2000 in order to avoid including very small systems which
were not considered in S230-BHX.

\begin{figure}
\centering
\includegraphics[width=0.4\textwidth]{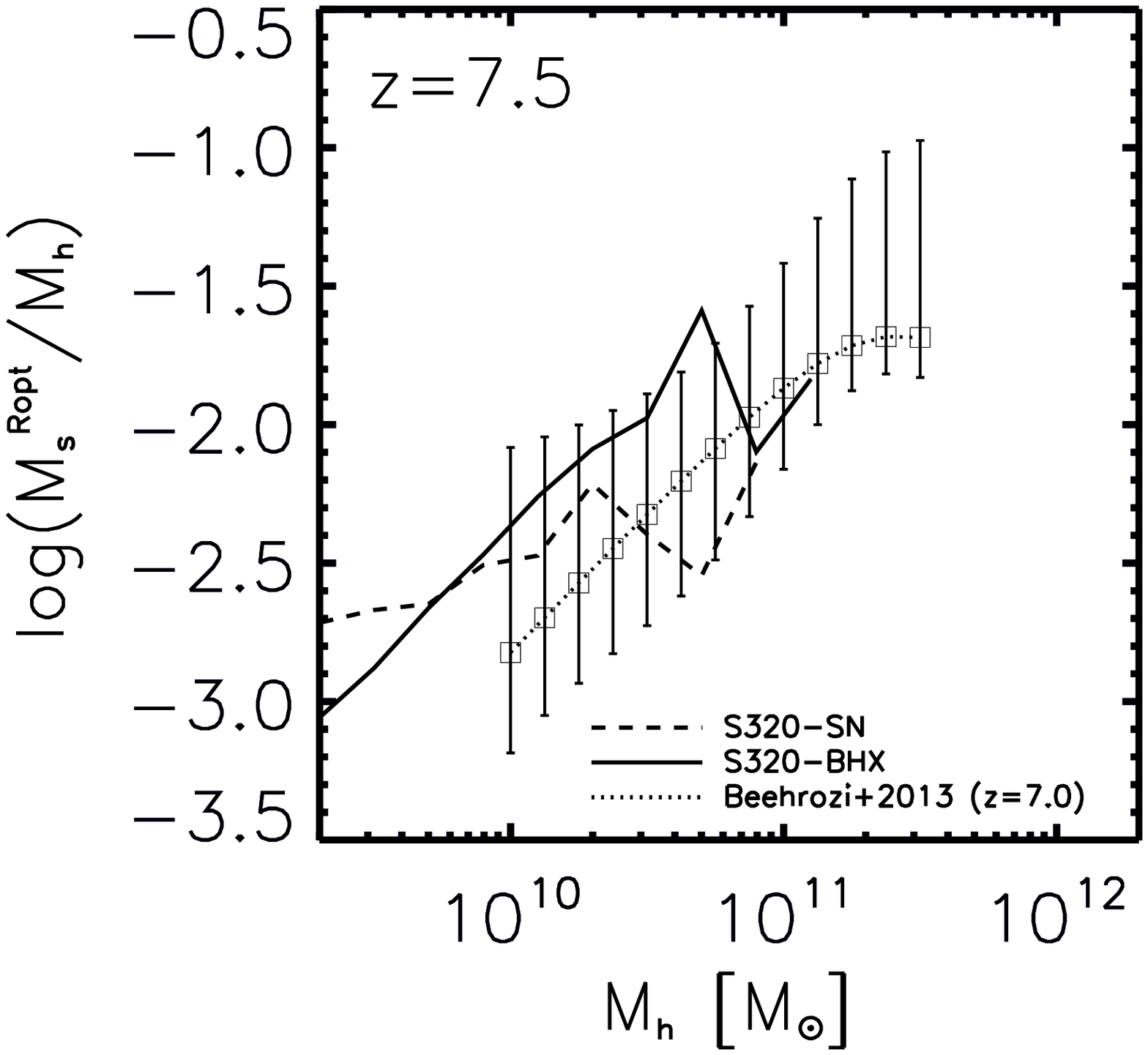}
\includegraphics[width=0.4\textwidth]{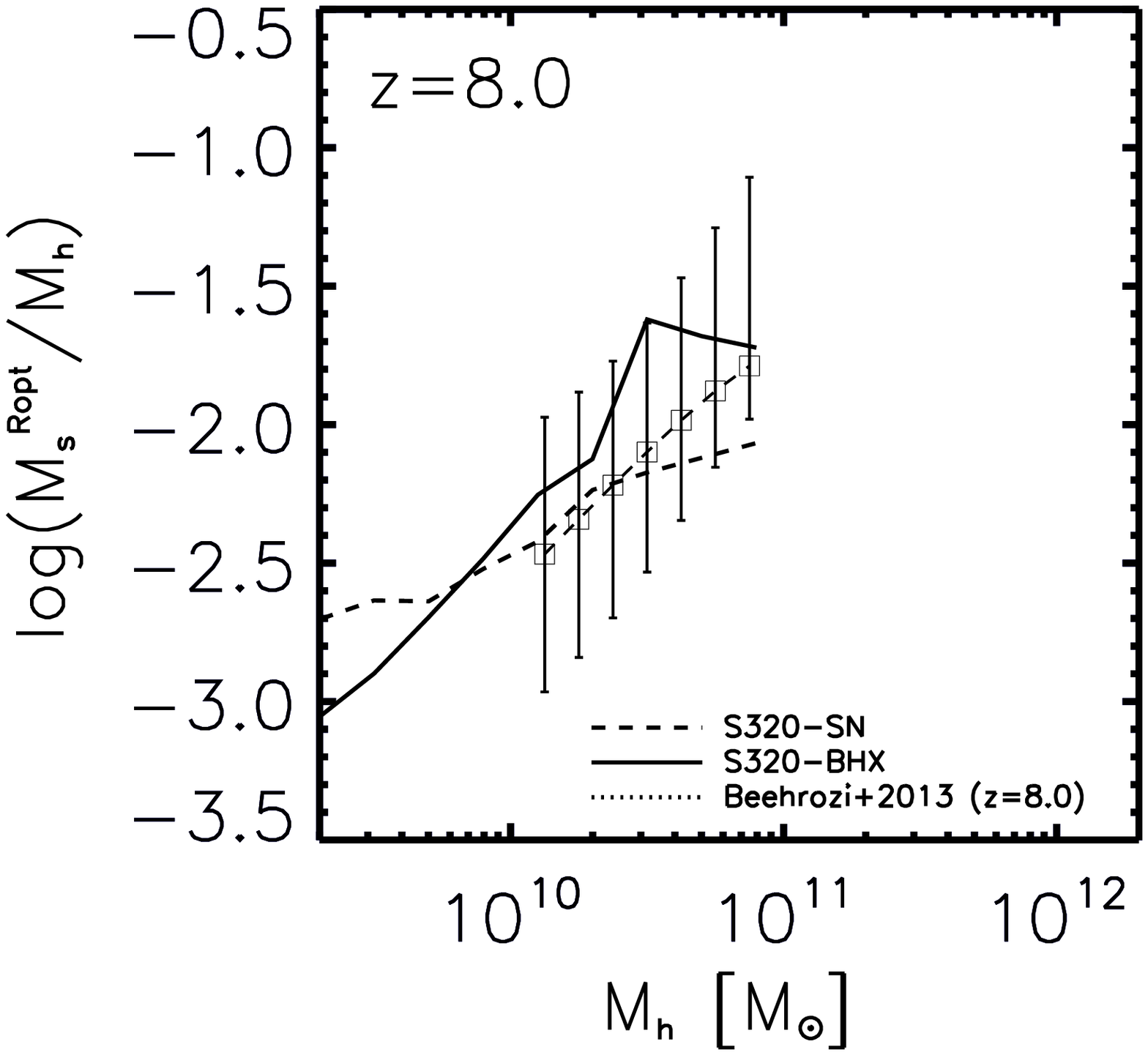}
\caption{Mean stellar-to-virial mass relation at $z =  7$ and $z = 8$, for
  simulated galaxies in S320-SN (dashed lines) and S320-BHX (solid lines).
The results of abundance matching technique from \citet{Behroozi2013}
are included (dotted line plus empty squares).  The main impact of BHX feedback occurs for $M_{\rm h} <10^{10} \, M_\odot$, in agreement with results found for
S230-BHX.}
\label{ms_mh_320}
\end{figure}

In Fig.~\ref{ms_mh_320} we show the stellar-to-virial mass relation in a 
similar fashion as in  Fig.~\ref{mh_ms}. The relations for galaxies in S320-BHX and S320-SN cross
each other at about $M_{\rm h} \sim 10^{10} \, M_\odot$ in agreement
with the trends found in the S230 runs. 
Even more,  the cumulative distribution of stellar mass also exhibits a
  similar trend to those found in S230 (Fig. ~\ref{histogramas}).
 And as a consequence, this agreement between the low and high resolution runs suggest that the results are
numerically robust.

\subsection{Contribution of X-ray photons} \label{phot_res}

As already discussed, the rate and X-ray luminosity of BHXs are expected to be
highly related to the metallicity of the progenitor stars as suggested by observational and theoretical studies.
 Due to the long-mean
free path of X-ray photons, BHXs could contribute  significantly to
the heating
and ionization of the IGM at early stages of galaxy evolution
\citep[e.g.][]{Mirabel2011,Jeon2013,Power2013,Knevitt2014}. \citet{Mirabel2011} show how 
the evolution of the temperature of the low-density neutral IGM due to
heating by X-rays from BHXs may depend on the  value of the parameter $f_{\rm X}$.
This parameter is defined as the ratio of the total X-ray luminosity $L_{\rm X}$
of a galaxy in the 2--10~keV band to its SFR, and its value can be estimated
from the expression

\begin{equation}
\label{fx}
f_{\rm X} = (3.5 \times 10^{40}\,  {\rm SFR})^{-1} L_{\rm X},
\end{equation}

\noindent
where $L_{\rm X}$ is measured in ${\rm erg \, s}^{-1}$ and the SFR in
$M_\odot\,{\rm yr}^{-1}$. An increase of $f_{\rm X}$ would cause the neutral IGM to be
heated earlier due to the presence of BHXs. This could limit the cold gas accretion in  dwarf galaxies at high
redshift.

Although the  detailed analysis of X-ray feedback of BHXs into the IGM is out of the
scope of this work, we estimate the value of $f_{\rm X}$ using our model for
extremely metal-poor BHXs in order to evaluate its compatibility with
expectations from other models. The emission of these BHXs should dominate the X-ray
luminosity of star-forming galaxies mainly at high redshift when most young stellar
populations were metal-poor. In any case, our predicted $f_{\rm X}$
represents a lower limit, resulting in a conservative contribution to
radiative feedback in the early  Universe.

Assuming that BHXs  radiate a similar amount of energy in
X-ray luminosity to that  injected in kinematic outflows (i.e.
$\epsilon_{\rm X} = E_{\rm BHX} = 10^{52}\, {\rm erg}$), and a mean
emission time of $\sim 3$ Myr,  the mean X-ray
luminosity per source as $L_{\rm X} \sim 10^{38}\,{\rm erg \, s}^{-1}$, as already mentioned in Section 2.2.
If the number of BHXs produced in a galaxy is $N_{\rm BHX}$, then the mean total X-ray
luminosity of a simulated galaxy originated from BHXs is $L_{\rm X}^{\rm G} = N_{\rm BHX}  L_{\rm X}$,
which can be easily computed at each time step of the
simulation. To make a proper comparison with \citet{Mirabel2011}, we use
the galaxy catalogue of S320-BHX at $z \sim  9$.
 These simulated galaxies have
${\rm SFR}$ and $L_{\rm X}^{\rm G}$ in the range $[0.01-20]\, M_{\odot} {\rm yr^{-1}}$ and
$[10^{39}-10^{41}]{\rm erg\, s}^{-1}$, respectively. Hence, the mean $f_{\rm X}$ value
obtained at $z \sim 9$ is $\langle f_{\rm X} \rangle = 0.65$.

Note that the total X-ray luminosities of simulated galaxies
do not exceed the X-ray luminosity limit of the deepest surveys \citep[4~Ms Chandra Deep Field-South ---CDF-S---,][]{Xue2011}.
At $z \sim 9$, the 2--10~keV rest
frame energy band corresponds roughly to the observed soft band
0.2--1~keV.  The flux limit of the 4~Ms CDF-S in the soft band is $9.1 \times 10^{-18} {\rm erg\,cm^{-2}\,s}^{-1}$, 
therefore the (k-corrected) luminosity limit at
$z \sim 9$ is $L_{\rm X, lim} \sim 4.6 \times 10^{42} \,{\rm erg \, s}^{-1}$.
Hence, our BHXs model does not contradict the upper limits imposed by current surveys.

The $\langle f_{\rm X} \rangle$ value obtained does not take into account the absorption of
X-ray photons before reaching the IGM.  
A rough estimation of the fraction of escaping photons  into the IGM
could be obtained by  adopting a power law spectral
energy distribution for the sources, with an index of $1.7$ in the
0.5--10~keV range \citep{Swartz2004}. 
We calculate this fraction for a typical halo of $M_{\rm h}=10^{8}\,
M_{\odot}$, assuming an exponential surface density  for the galaxy.
We assume that the ratio of the baryonic-to-dark-matter mass is
$m_{\rm d} = 0.17$, the spin parameter is $\lambda \sim 0.03$ and
$j_{\rm d} / m_{\rm d} = 1$, where $j_{\rm d}$ is the ratio of the
angular momentum of baryons to that of the halo \citep[][]{Mo1998, Fernandez2011}. 
%These values were selected following previous results \citep{Fernandez2011}.
Considering that X-ray photons would ionize both hydrogen and helium in the
IGM, we use the cross sections of \citet{Kuhlen2005}.
The estimated escape fraction is then $f_{\rm esc} \sim 0.53$.  Therefore, the
effective $\langle f_{\rm X} \rangle$ 
would be $\langle f_{\rm X} \rangle_{\rm eff} \sim  0.34$ at redshift $z \sim 9$.
This value  is lower than that proposed by
\citet{Mirabel2011} where  $f_{\rm esc} \sim 1.0$ is assumed. Our simple estimation
 implies that the generated X-ray photons could barely heat the
temperature of the IGM above $10^{3}\,{\rm K}$.
However, our estimations are too crude to  draw  more
robust conclusions on this aspect.  This effect can be studied with radiative transfer models in detail  
\citep[e.g.][]{Knevitt2014}.

\section{Conclusions} \label{concl}

In this work, we explore the effects of HMXBs composed by an accreting black hole 
 on galaxy evolution \citep{Justham2012}.
Following previous results, our model assumes that the rate, X-ray luminosity and
outflow energy of BHXs
increase for low metallicity progenitors 
\citep{Dray2006,Linden2010}.
The BHX feedback model is grafted into a version of {\small P-GADGET-3}
which includes SN feedback and chemical evolution
\citep{Scannapieco2006}, so that its effects can be studied as the structure forms and evolves.

The estimated kinetic energy released by BHXs is assumed to be
efficiently thermalized and hence,  is pumped into the cold gaseous phase in
the surrounding region  of the
sources. This extra source of energy has an impact on  the properties of the ISM  at early
stages of evolution and in low-velocity haloes. Encouragingly, our results agree with
those reported by \citep{Justham2012} where a semi-analytical model
was used to implement a scheme based on similar hypotheses.

Our results can be summarized as follows:

\begin{itemize}

\item 
Following observations which
indicate that  the kinetic energy deposited by these sources might
be comparable to that radiated in the X-ray bands (during $\sim 3$
Myr), we estimate that each BHX event could inject $\sim 10^{52}$ erg in form of kinetic
energy in the ISM. With these hypotheses and in order to reproduce the observed cSFR,
 our BHX model requires that a fraction of $\sim 20 \%$
of  BHs with metallicity in the range $Z_{\rm EMP} = [0,10^{-4}]$ should end up as X-ray binary systems. 
We tested combinations of larger and lower released kinetic
energy and BHX fractions finding that they produce  too strong or too
weak effects, resulting in  an inadequate regulation of the star formation
at high redshift.

\item  Our model predicts  BHX feedback to affect more strongly 
 galaxies with  stellar masses in the range
$10^7$--$10^8\, M_\odot$ 
which  inhabit haloes with virial velocities smaller than $V_{\rm vir} \sim 40\, {\rm km \,
  s}^{-1}$.  Larger
haloes would experience a negligible impact due to their larger 
potential wells but
they  could be  indirectly affected  by the accretion of smaller
haloes.
These results are shown to be robust against numerical resolution.

\item The  energy injected 
by BHXs helps to regulate the star formation activity in low-mass
haloes, decreasing the fraction of stars formed at very high redshift. 
As a  consequence, there is more gas available for star
formation at later times. 
Therefore, when BHX feedback is included, the cSFR reproduces closely
observational results  \citep{Behroozi2013}.  In this
simulated volume, we found the BHX energy represents a $\sim 30\%$ of
total feedback energy released into the ISM at $z\sim7$,
and a $10\%$  at $z\sim 4$.

\item When BHX feedback is included the number of galaxies as a function of stellar masses shows
 a larger evolution with redshift (for $z > 4$) compared to the run with only SN
 feedback turned on. The variation of the evolution is driven by the
 low-mass haloes which are more significantly affected by BHX
 feedback, and hence, its characteristics could
be an observational test to probe the action of this feedback.

\item The stellar-to-virial mass relation for $z\sim  7$ is in better
  agreement to that predicted from abundance matching  \citep{Moster2013,Behroozi2013}. For
 lower redshift, there is still  an excess of stars,
although the slope of the relation is more comparable to the predicted trends.  

\item Simulated galaxies with BHX+SN feedbacks exhibit hot ISM
  at very high redshift but their mean  temperature is smaller
  than the virial temperature of the haloes. If the BHX feedback is
  switched off, the resulting ISM are colder and  the fraction of new born
  stars higher in low mass haloes.  As a consequence the number of SNe is enough to blow
  part of the ISMs,  contributing to  building  up hotter CGMs.  In order
  to prevent this, the SN feedback should be weaken but this would
  lead to a further  increase of the star formation, producing a larger
  stellar fraction.

\end{itemize}

Hence, our results support previous claims that the BHXs could be an important source
of feedback in early stages of galaxy evolution by regulating the
star formation in haloes with $V_{\rm vir} \leq 40\, {\rm km \, s}^{-1}$.  It goes in the same direction as the
so-called 'enhanced' feedback although BHX feedback would have
significant impact
only in the very early Universe.

\section*{Acknowledgements}
We would like to thank the referee, Dr Graham Wynn, for his careful
reading and comments which helped to improve the paper, and Laura Sales for
useful discussions.
Simulations are part of the Fenix Project and  have been run in  Hal Cluster of the Universidad Nacional
de C\'ordoba, AlphaCrucis  of
IAG-USP (Brasil) and Barcelona Supercomputer Center. We acknowledged the use of   Fenix Cluster of
Institute for Astronomy and Space Physics.
This work has been partially supported by PICT
Raices 2011/959 of Ministery of Science (Argentina) and Proyecto
Interno of Universidad Andres Bello (Chile).

\bibliographystyle{mn2e}

\def\apj{ApJ}
\def\apjs{ApJS}
\def\apjl{ApJ}
\def\aj{AJ}
\def\mnras{MNRAS}
\def\aa{A\&A}
\def\nat{Nature}
\def\araa{ARA\&A}
\def\aap{A\&A}
\def\pasp{PASP}
\def\nar{New Astronomy Reviews}
\bibliography{ArtaleMC_feedback_REVFINAL}

\IfFileExists{\jobname.bbl}{}
{\typeout{}
\typeout{****************************************************}
\typeout{****************************************************}
\typeout{** Please run "bibtex \jobname" to optain}
\typeout{** the bibliography and then re-run LaTeX}
\typeout{** twice to fix the references!}
\typeout{****************************************************}
\typeout{****************************************************}
\typeout{}
}

\end{document}